\documentstyle[11pt]{article}
\advance \topmargin by -\headheight
\advance \topmargin by -\headsep

\evensidemargin \oddsidemargin
\title{
       Approximate Lie Group Analysis of Finite--difference \\
       Equations\\
      }

\author{\large A.M.Latypov \\
               Fluid Dynamics Research Institute \\
               and Department of Mathematics and Statistics, \\
               University of Windsor, \\
               Windsor, Ontario N9B 3P4, \\
               CANADA. \\
               \mbox{ } \\
               Internet e-mail: latypov@server.uwindsor.ca  }

\date{November 6, 1995}

\begin{document}

\newtheorem{definition}{Definition}
\newtheorem{theorem}{Theorem}
\newtheorem{lemma}{Lemma}
\newcommand{\down}[2]{\begin{array}[t]{c}{#1}\\{^{^#2}} \end{array}}
\newcommand{\mf}{\mathop{\rm{\bf f}}}
\newcommand{\mX}{\mathop{\rm{X}}}
\newcommand{\mF}{\mathop{\rm{\bf F}}}
\newcommand{\mz}{\mathop{\rm{\bf z}}}
\newcommand{\mbxi}{\mathop{\rm{\mbox{\boldmath $\xi$}}}}
\newcommand{\mxi}{\mathop{\rm{\mbox{$\xi$}}}}
\newcommand{\mvarphi}{\mathop{\rm{\mbox{$\varphi$}}}}
\newcommand{\msigma}{\mathop{\rm{\mbox{$\sigma$}}}}
\newcommand{\R}{{\sf R\hspace*{-1.6ex}\rule{0.15ex}%
{1.4ex}\hspace*{0.9ex}}}

\bibliographystyle{plain}

\maketitle

\begin{abstract}
Approximate group analysis technique, that is, the technique combining
the methodology of group analysis and theory of small perturbations,
is applied to finite--difference equations approximating ordinary
differential equations.

Finite--difference equations are viewed as a system of algebraic
equations with a small parameter, introduced through the definitions
of finite--difference derivatives. It is shown that application of
the approximate invariance criterion to this algebraic system results in
relations that can be viewed as prolongation formulae and the invariance
criterion for the differential approximation of these finite--difference
equations.

This allows us to study the group properties of the finite--difference
equations by analyzing the group properties of their differential
approximations, which are the differential equations with a small
parameter.

In particular, the question of whether the group, admitted by the original
differential equation, can be corrected by adding the first--order
perturbation to it, so that the resulting group with a small parameter
is approximately admitted by the finite--difference approximation,
is studied. It is shown by examples that, for a given differential
equation, its finite--difference approximation and the group, such a
correction may not always be possible. It is also demonstrated that the
finite--difference approximation can be modified in such a way that
the correction becomes possible.

\end{abstract}



\section{Introduction}
An invariance of the differential equations with respect to
one--parameter groups of transformations in the space of
independent and dependent variables carries important
information about the fundamental properties, such as
conservation laws,  of a physical system that these
equations may describe.

Knowledge of the groups admitted by the system of differential
equations may also allow one to reduce the order of the system,
to find important particular solutions, or to produce new solutions
from a single solution which is known \cite{OLVER}.
Preserving the group properties of differential equations
in their discretizations seems to be a natural requirement
which would ensure that the fundamental properties of the
approximated differential equation are preserved in the discrete
model.

A step in the direction of analyzing the properties of purely
difference equations was made by Maeda \cite{MAEDA}, where the questions
of linearization and reduction of order in ordinary difference
equations were treated. Levi and Winternitz \cite{WINTERNITZ}
applied a Lie groups technique
to differential--difference equations arising in study
of lattices. Knowledge of these group properties allowed the authors
to obtain some non--trivial invariant solutions.

Dorodnitsyn \cite{DORODNITSYN} developed a general theory of
transformation groups
in the spaces of discrete variables. By properly
constructing the space of discrete variables and the finite--difference
analogues of the differential operators, he demonstrated that
the  methodology results in the prolongation formulae for
the finite--difference derivatives, which are similar to those formulae
in the continuous case. This similarity allowed the author to extend some
of the group--analytical methods to the finite--difference equations.
In particular, the difference analogy of the Noether theorem is proven.

Unfortunately, in general, replacing the system of differential
equations with their finite--difference, finite volume or
finite element discretizations  introduces noninvariance
into the discretization. This noninvariance stems from the fact
that performing the discretization requires utilizing a discrete set
of nodes in the space of independent variables -- a computational grid.
Formally this implies that the group of transformations acting on the
discrete variables needs to leave invariant not only the approximating
finite--difference system, but also a set of algebraic equations defining
the grid \cite{DORODNITSYN}. Although some of the important groups
of transformations arising in applications do satisfy this property,
generally speaking, this requirement results in the discrete system
losing many of the groups of the original differential equation.

The natural way to avoid this restrictiveness of the above approach
is to weaken the requirement of invariancy of the finite--difference
system. One of the possibilities, suggested and studied by Shokin
\cite{SHOKIN}, is to consider the {\it differential approximation}
of the finite--difference
system. By formally applying a Taylor series expansion to the
finite--difference approximation, one can obtain a differential
equation which can be treated as a perturbation of the original
differential equation due to discretization. The perturbation
terms, added to the original equation, depend on the step sizes
of the grid.

In \cite{SHOKIN} the general theory of differential approximations
of the discrete models is developed and group properties of the
differential approximations are studied. Following this methodology
one can find the groups admitted by the first differential
approximations and deduce some important properties of the discrete
model from analysis of these groups.

Despite the simplification introduced by  considering the first
differential approximations instead of the system of algebraic
equations, the approach \cite{SHOKIN} still requires answering some
important questions.

The step sizes of the grid participating in the differential
approximation are the parameters which  need to be transformed
when a group of transformations acts on the independent variable.
Since the step size is defined as the difference between the values
of the independent coordinate at the end points of the mesh interval,
it can easily be seen that the
{\em defining equations} \cite{OLVER} for the general symmetry
group of the given system will no longer be local in the sense
that they will contain not only the derivatives of the infinitesimal
operator's components, but also the values of these components
at the neighboring nodes. Although for many important symmetry groups
this non--locality can be avoided \cite{SHOKIN}, this fact makes
finding the general symmetry group as hard as solving the
finite--difference equation in closed form.

Another important question related to the approach \cite{SHOKIN}
follows from noticing that, in general, the symmetry group admitted
by the differential approximation of the difference equation does not
leave the original difference equation invariant. Therefore,
further formalization is needed in order to study the properties
of the discrete model through analyzing the groups of transformations
admitted by the differential approximation.

In this work
the recently developed technique of approximate group analysis
due to Baikov, Gazizov and Ibragimov \cite{BAIKOV:1} is used
as a basis for such a formalization.
The discrete model is viewed as the system of algebraic
equations with a small parameter introduced through the definitions
of the grid step sizes and finite--difference derivatives. The groups
of transformations leaving this system of algebraic equations together
with the definitions of step sizes and the difference derivatives
approximately invariant \cite{BAIKOV:1} are then sought.
It is demonstrated that the groups of transformations satisfying
these properties can also be found by considering the differential
approximation of the discrete model, provided that the action
of the group on the step sizes
is extended in a special way. This ``prolongation'' of the group
to the step sizes is local and therefore it is easier to solve the
determining system analytically.

\section{Approximate symmetries}
In order to make the presentation complete and to introduce some
important concepts to be used in the following, the basic notions
of approximate symmetries are outlined in this Section
following \cite{BAIKOV:1}.
\subsection{Approximate one--parameter groups of transformations}
Let ${\bf z} = (z_1,...,z_N) \in \R^{\;\;N}$ denote an independent
variable, $\varepsilon$ be a small parameter, and $p$ be a fixed
natural number.

Consider the following one--parameter family of approximate
transformations acting in $\R^{\;\;N}$:
\begin{eqnarray}
  {\bf z}^\prime = {\bf f}({\bf z}, \varepsilon, a) +
o({\varepsilon}^p),\label{eq:transformation}
\end{eqnarray}
where ${\bf f} = (f_1,...,f_N)$ and $a$ is a scalar parameter of this family
of transformations.
\begin{definition}
We say that transformations (\ref{eq:transformation}) form an approximate
one--parameter group (a.o.p.g.) with respect to the parameter $a$ if
\begin{eqnarray}
  {\bf f}({\bf z},\varepsilon,0) = {\bf z} + o({\varepsilon}^p),
\label{eq:initial} \\
  {\bf f}({\bf f}({\bf z}, \varepsilon, a),\varepsilon,b) =
  {\bf f}({\bf z}, \varepsilon, a + b)
   + o({\varepsilon}^p)                                   \label{eq:group}
\end{eqnarray}
and also $a=0$ if $\;\;{\bf f}({\bf z},\varepsilon,a) = {\bf z} +
o({\varepsilon}^p)$
is true for all ${\bf z}$.
\end{definition}
\begin{definition}
An approximate infinitesimal operator corresponding to an a.o.p.g.
(\ref{eq:transformation}) is
\begin{eqnarray}
  X = \mbox{\boldmath $\xi$}^T \partial_{\bf z} + o({\varepsilon}^p),
\label{eq:operator}
\end{eqnarray}
where
  $\partial_{\bf z} = (\partial_{z_1},...,\partial_{z_N})^T$,
  \(\mbox{\boldmath $\xi$} = ({\xi}_1,...,{\xi}_N)^T\) and
  ${\xi}_{i}({\bf z}, \varepsilon) = \frac{\partial}{\partial a} {f_i}
{\mid}_{a=0}$,
  $i=1,...,N$.
\end{definition}

The following analogue of the Lie theorem \cite{OLVER} for a.o.p.g.
establishes a relation between an a.o.p.g. (\ref{eq:transformation})
and its approximate infinitesimal operator (\ref{eq:operator}).
The proof can be found in \cite{BAIKOV:1}.
\begin{theorem}[Lie theorem for a.o.p.g.]
 \label{theorem:Lie}
If the transformation (\ref{eq:transformation}) forms an a.o.p.g.
with an approximate infinitesimal operator (\ref{eq:operator}), then
the function ${\bf f}({\bf z},\varepsilon,a)$ satisfies the equation
\begin{eqnarray}
\frac{\partial}{\partial a}  {\bf f}({\bf z},\varepsilon,a) =
 \mbox{\boldmath $\xi$}({\bf f}({\bf z},\varepsilon,a),\varepsilon)
 + o({\varepsilon}^p).                                    \label{eq:Lie}
\end{eqnarray}
Conversely, for arbitrary smooth function, the solution
(\ref{eq:transformation}) of the Cauchy problem
\begin{eqnarray*}
\frac{d {\bf z}^\prime}{d a} = \mbox{\boldmath $\xi$}({\bf
z}^\prime,\varepsilon)
 + o({\varepsilon}^p),     \\
{\bf z}^\prime |_{a=0} = {\bf z} + o({\varepsilon}^p)
\end{eqnarray*}
yields an a.o.p.g. with group parameter $a$.
\end{theorem}
\mbox{ } \\
\mbox{ } \\
{\bf Example} \cite{BAIKOV:2}, \cite{BAIKOV:1}. Consider the case $p = 1$ .
Assuming that the function ${\bf f}({\bf z}, \varepsilon, a)$ is analytic,
write the a.o.p.g. (\ref{eq:transformation}) as:
\begin{eqnarray}
   {\bf z}^\prime =   \mf_0({\bf z},a)+\varepsilon \mf_1({\bf z},a)+
                  o(\varepsilon)                \label{eq:f_1st_order}
\end{eqnarray}
An approximate infinitesimal operator can then be written as
\begin{eqnarray}
 X = \mX_0 + \varepsilon \mX_1+o(\varepsilon) =
                 {\mbxi_0}^T({\bf z}) \partial_{\bf z} +
     \varepsilon {\mbxi_1}^T({\bf z}) \partial_{\bf z} +
   o(\varepsilon),                              \label{eq:X_1st_order}
\end{eqnarray}
where
\begin{eqnarray*}
  {\mbxi_0}({\bf z})=
  \left.\frac{\partial}{\partial a} \mf_0({\bf z},a)\right|_{a=0},
\;\;\;\;\;
  {\mbxi_1}({\bf z})=
  \left.\frac{\partial}{\partial a} \mf_1({\bf z},a)\right|_{a=0},
\end{eqnarray*}
Using (\ref{eq:f_1st_order}) and (\ref{eq:X_1st_order}) in
Theorem~\ref{theorem:Lie} we find that ${\displaystyle \mf_0}$ and
${\displaystyle \mf_1}$ satisfy the equations
\begin{eqnarray*}
 \frac{\partial}{\partial a} \mf_0 &=& \mbxi_0 (\mf_0), \\
 \frac{\partial}{\partial a} \mf_1 &=& \mbxi_1 (\mf_0) + {\mbxi_0}^\prime
(\mf_0) \mf_1,
\end{eqnarray*}
where ${\displaystyle {\mbxi_0}^\prime = \frac{\partial }{\partial {\bf z}}
\mbxi_0}$.
Conversely, the functions ${\displaystyle \mf_0}$ and ${\displaystyle \mf_1}$
can be found from the
approximate infinitesimal operator (\ref{eq:X_1st_order}) by solving
the following Cauchy problem:
\begin{eqnarray} \label{eq:Cauchy}
 \frac{d}{d a} \mf_0 &=& \mbxi_0 (\mf_0), \label{eq:Cauchy1} \\
 \frac{d}{d a} \mf_1 &=& \mbxi_1 (\mf_0) +
                        {\mbxi_0}^\prime (\mf_0) \mf_1,  \label{eq:Cauchy2}\\
 \left.\mf_0\right|_{a=0} = {\bf z},\;\;& &\;\;\left.\mf_1\right|_{a=0} = {\bf
0}.  \label{eq:Cauchy3}
\end{eqnarray}

\subsection{Approximate invariance}
Let $F_i({\bf z}, \varepsilon)$, $i=1,...,M$ be  given scalar--valued
functions and
\begin{eqnarray}
  {\bf F({\bf z}, \varepsilon)}
  =(F_1({\bf z}, \varepsilon),...,F_M({\bf z}, \varepsilon))^T. \label{eq:RHS}
\end{eqnarray}
\begin{definition}
The equation
\begin{eqnarray}
  {\bf F}({\bf z}, \varepsilon) = o({\varepsilon}^p)
\label{eq:approximate_equation}
\end{eqnarray}
is said to be invariant with respect to an a.o.p.g. (\ref{eq:transformation}),
if, for all ${\bf z}$ satisfying (\ref{eq:approximate_equation}), we have
\begin{eqnarray}
  {\bf F}({\bf f}({\bf z}, \varepsilon, a), \varepsilon) =
  o({\varepsilon}^p).  \label{eq:invariance}
\end{eqnarray}
\end{definition}

The following criterion of invariance has been proven in \cite{BAIKOV:1}.
\begin{theorem}[Invariance criterion]
                    \label{theorem:infinitesimal_criterion}
Let the function (\ref{eq:RHS}) be analytic with respect to the variables
${\bf z}$  and $\varepsilon$ and also
\begin{eqnarray}
 rank \left. \left(\frac{\partial {\bf F}}{\partial {\bf z}}({\bf z}, 0)\right)
 \right|_{{\bf F}({\bf z}, 0)=0} = M.     \label{eq:nondegeneracy}
\end{eqnarray}
Then the approximate equation (\ref{eq:approximate_equation}) is invariant
with respect to an a.o.p.g. (\ref{eq:transformation}) with an approximate
infinitesimal operator (\ref{eq:operator}) if and only if
\begin{eqnarray*}
   X {\bf F}({\bf z}, \varepsilon) |_{(\ref{eq:approximate_equation})}
   = o({\varepsilon}^p).
\end{eqnarray*}
\end{theorem}
\mbox{ }\\
\\
{\bf Example} \cite{BAIKOV:2}, \cite{BAIKOV:1}. Let us consider the
case $p = 1$ and
obtain the necessary and sufficient condition under which the equation
\begin{eqnarray}
 {\bf F}({\bf z},\varepsilon) \equiv
   \mF_0({\bf z}) + \varepsilon \mF_1({\bf z}) = o(\varepsilon),
                      \label{eq:F_1st_order}
\end{eqnarray}
where ${\displaystyle \mF_0}$ and ${\displaystyle \mF_1}$ are both analytic
functions, is invariant with respect to the a.o.p.g. (\ref{eq:f_1st_order}),
(\ref{eq:X_1st_order}). First notice that substituting
(\ref{eq:f_1st_order}) into (\ref{eq:invariance}), expanding
the result in powers of $\varepsilon$ and taking into account
analyticity of the involved functions results in the
following set of ``exact'' equations, which is equivalent
to the invariance condition (\ref{eq:invariance}) with $p=1$:
\begin{eqnarray*}
 \mF_0(\mf_0({\bf z},a)) &=& {\bf 0}, \\
 \mF_1(\mf_0({\bf z},a)) +
 {\mF_0}^\prime(\mf_0({\bf z},a)) \mf_1({\bf z},a) &=& {\bf 0},
\end{eqnarray*}
where
${\displaystyle {\mF_0}^\prime = \frac{\partial}{\partial {\bf z}}{\mF_0} }$.

Suppose that ${\displaystyle \mF_0}$ satisfies
the nondegeneracy condition following from (\ref{eq:nondegeneracy}).
Then the infinitesimal criterion of invariance can be written,
according to Theorem~\ref{theorem:infinitesimal_criterion},
as follows:
\begin{eqnarray}
   X  \left.
         \left(
            \mF_0({\bf z}) + \varepsilon \mF_1({\bf z})
         \right)
       \right|_{(\ref{eq:F_1st_order})}
 = o(\varepsilon),
                            \label{eq:invariance_1st_order}
\end{eqnarray}
where $X$ is given by (\ref{eq:X_1st_order}).
Following \cite{BAIKOV:2}, \cite{BAIKOV:1}, we substitute
${\bf z}={\displaystyle \mz_0 + \varepsilon \mz_1} + o(\varepsilon)$ into
(\ref{eq:invariance_1st_order}) in order to obtain the equivalent
set of exact equalities:
\begin{eqnarray}
 \mX_0 \mF_0(\mz_0) &=& 0, \label{eq:determining_1st_order_1}\\
 \mX_1 \mF_0(\mz_0) + \mX_0 \mF_1(\mz_0)
                    + \mz_1 \partial_{\bf z} (\mX_0 \mF_0(\mz_0)) &=& 0.
                           \label{eq:determining_1st_order_2}
\end{eqnarray}
The above two equalities must be satisfied when
\begin{eqnarray}
  \mF_0(\mz_0) &=& {\bf 0},
                       \label{eq:F_1st_order_1}\\
  \mF_1(\mz_0) + \mz_1 \partial_{\bf z} \mF_0 (\mz_0) &=& {\bf 0},
                       \label{eq:F_1st_order_2}
\end{eqnarray}
as follows from (\ref{eq:F_1st_order}).

In the following we further restrict our consideration to the case $M=1$,
that is ${\bf F} = F_1({\bf z}, \varepsilon)$.
In practice, as suggested in \cite{BAIKOV:2}, \cite{BAIKOV:1},
it is convenient to get rid of ${\displaystyle \mz_1}$
in the above relations by employing the following fact.
It can easily be seen from (\ref{eq:determining_1st_order_1})
and (\ref{eq:F_1st_order_1}) that ${\displaystyle \mX_0}$
is an infinitesimal operator corresponding to the exact
transformation group which leaves equation
(\ref{eq:determining_1st_order_1}) invariant. Furthermore,
as follows from Proposition 2.10 from \cite[p.84]{OLVER},
the left--hand side of (\ref{eq:determining_1st_order_1})
can be written as follows:
\begin{eqnarray}
   \mX_0 \;\; \mF_0 (\mz_0) = \lambda(\mz_0) \mF_0 (\mz_0),
                     \label{eq:lambda}
\end{eqnarray}
where $\lambda(\mz_0)$ is some smooth function. Then, using
(\ref{eq:F_1st_order_2}) and substituting (\ref{eq:lambda})
into (\ref{eq:determining_1st_order_2}), we rewrite the
determining system (\ref{eq:determining_1st_order_1}),
(\ref{eq:determining_1st_order_2}) together with the
relations (\ref{eq:F_1st_order_1}) and (\ref{eq:F_1st_order_2})
as follows:
\begin{eqnarray}
   \mX_0 \mF_0 (\mz)                    &=& \lambda(\mz) \mF_0 (\mz),
                  \label{eq:determining_1st_order_simplified_1}       \\
   \mX_1 \mF_0 (\mz) + \mX_0 \mF_1(\mz) &=& \lambda(\mz) \mF_1 (\mz).
                  \label{eq:determining_1st_order_simplified_2}
\end{eqnarray}
If the invariance group for the (``non--perturbed'') equation
${\displaystyle \mF_0 = {\bf 0}}$ is known, the relation
(\ref{eq:determining_1st_order_simplified_1}) can be used to
determine the smooth function $\lambda(\mz)$. This function
is then substituted into (\ref{eq:determining_1st_order_simplified_2}),
which must be satisfied by virtue of (\ref{eq:F_1st_order_1}).
Solving (\ref{eq:determining_1st_order_simplified_2}) together
with (\ref{eq:F_1st_order_1}), one finds the ``perturbation''
${\displaystyle \mX_1}$ of the infinitesimal operator.
\mbox{ }\\
\\
{\bf Example} \cite{BAIKOV:1}. Letting $p=1$, $N=2$,
 ${\bf z}=(z_1,z_2)^T$, equation
(\ref{eq:F_1st_order}) reads
\begin{eqnarray*}
 {\bf F}({\bf z},\varepsilon)
  \equiv \mF_0({\bf z}) + \varepsilon \mF_1({\bf z})
  \equiv z_2^2 - 1 + \varepsilon( z_2^2 \ln z_2 - z_1^2)  = o(\varepsilon),
\end{eqnarray*}
The above equation can be considered as a small perturbation
of the equation ${\displaystyle \mF_0({\bf z}) = 0}$. The latter equation
obviously admits the group ${\displaystyle \mX_0 = \partial_{z_1}}$.
One may consider a small perturbation of this group given by
a.o.p.g.
 ${\displaystyle \mX = \mX_0 + \varepsilon \mX_1 + o(\varepsilon)}$ and
establish the condition on ${\displaystyle \mX_1}$
ensuring that the a.o.p.g. leaves the equation
invariant.

First notice that (\ref{eq:determining_1st_order_simplified_1})
results in ${\displaystyle\lambda(\mz) \equiv 0}$. Using it in
(\ref{eq:determining_1st_order_simplified_2}), one arrives to
the condition
\begin{eqnarray*}
 ({\mbxi_1}\mbox{}_2({\bf z}))^2 = (z_1)^2,
\end{eqnarray*}
which needs to be satisfied when $(z_2)^2 = 1$.

One a.o.p.g. satisfying this condition is
$X=(1+\varepsilon z_1^2)\partial_{z_1}+\varepsilon z_1 z_2 \partial_{z_2}
+o(\varepsilon)$.
Solving the Cauchy problem (\ref{eq:Cauchy1})--(\ref{eq:Cauchy3}),
one can find
the transformations corresponding to this a.o.p.g.:
\begin{eqnarray*}
  z_1^{\prime} &=& z_1 + a +
    \varepsilon (a z_1^2 + a^2 z_1 + \frac{a^3}{3}) + o(\varepsilon), \\
  z_2^{\prime} &=& z_2 +
    \varepsilon (a z_1 z_2 +     \frac{a^2 z_2}{2}) + o(\varepsilon).
\end{eqnarray*}

\section{Approximate symmetries of finite--difference equations}

\subsection{Finite-difference approximation as a system of equations
            with a small parameter}
Consider an ordinary differential equation of order $n$:
\begin{eqnarray}
  F(x,y,y^{(1)},...,y^{(n)})=0,\label{eq:ODE1}  \\
  y=y(x),\;\;\;\;x \in (a, b). \nonumber
\end{eqnarray}
Introduce the {\em grid} as the set of points $x_\alpha \in (a,b)$, $\alpha =
1,...,N$,
and the {\em grid function} $u_\alpha$,
 $\alpha = 1,...,N$ defined at these points and to be used to approximate
the values of $y=y(x)$ at the nodes of the grid.

Consider also the set of algebraic equations with respect to the
unknowns $x_\alpha$ and $u_\alpha$:
\begin{eqnarray}
  L_\alpha(x_1,...,x_N,u_1,...,u_N)=0,\;\;\;\;\alpha = 1,...,N. \label{eq:FDE1}
\end{eqnarray}
In the cases of interest to our work, this set of equations
{\em approximates} the differential equation (\ref{eq:ODE1}) in the
sense defined below.

In order to introduce a small parameter into (\ref{eq:FDE1}),
extend the system of equations (\ref{eq:FDE1}) by introducing the
new unknowns and adding the following equations to it:
\begin{eqnarray}
  x_{\alpha+1}-x_{\alpha} &=& \varepsilon h_\alpha + o(\varepsilon^p),
                           \label{eq:step_size} \\
  u_{\alpha+1}^{(J)}-u_{\alpha}^{(J)} &=&
  \sum_{k=1}^{p} \frac{\varepsilon^k}{k!} h_\alpha^k u_{\alpha}^{(J+k)}
  + o(\varepsilon^p), \label{eq:fd_derivatives} \\
  u_{\alpha^\prime}^{(0)} &=& u_{\alpha^\prime}, \label{eq:zero_derivative}\\
  \alpha = 1,...,N-1, \;\;\;\; \alpha^\prime &=& 1,...,N, \;\;\;\;
  p \geq n,
  \;\;\;\; J=0,...,p, \nonumber
\end{eqnarray}
where $\varepsilon$ is a small parameter and the added unknowns,
 $h_\alpha$ and $u^{(J)}_\alpha$, ($J=0,...,2p$)
will be referred to as,
respectively, {\em step sizes} and
{\em finite--difference derivatives of $J$-th order}.

Although this system is comprised of algebraic equations,
equations (\ref{eq:fd_derivatives}) can be viewed as a result
of a formal expansion in the Taylor series at the points $x=x_\alpha$,
which is used in order to obtain the values of $u_{\alpha+\beta}^{(J)}$,
where $\beta = 1$.
The following two Lemmas demonstrate that, by virtue of the equations
(\ref{eq:step_size}), (\ref{eq:fd_derivatives}) and
(\ref{eq:zero_derivative}), this analogy with the Taylor expansion
can be extended to the cases when $x_\alpha$ and $x_{\alpha + \beta}$
are two arbitrary nodes of the grid, and that it can be extended to
the smooth functions of the variables $x_\alpha$ and $u_\alpha$.

\begin{lemma} \label{lemma:expansion}
  If $1 \leq \alpha \leq N$, $\beta \neq 0$, $1 \leq \alpha + \beta \leq N$ and
  $0 \leq J \leq p-1$,
  then, by virtue of (\ref{eq:step_size}), (\ref{eq:fd_derivatives}) and
  (\ref{eq:zero_derivative})
  \begin{eqnarray}
    u_{\alpha+\beta}^{(J)}-u_{\alpha}^{(J)} =
    \sum_{k=1}^{p-J} \frac{\varepsilon^k}{k!}
    h_{\alpha+\beta,\alpha}^k u_{\alpha}^{(J+k)}
    + o(\varepsilon^{p-J}),                    \label{eq:lemma_statement}
  \end{eqnarray}
  where
  \begin{eqnarray*}
    h_{\alpha+\beta,\alpha} =
    \left \{
        \begin{array}{ll}
         {
          \displaystyle
          \sum_{i=\alpha}^{\alpha+\beta-1}
         } h_i & \mbox{if $\beta \geq  1$} \\
         {
          \displaystyle
         -\sum_{i=\alpha+\beta}^{\alpha-1}
         } h_i & \mbox{if $\beta \leq -1$}
        \end{array}
    \right.
  \end{eqnarray*}
\end{lemma}
{\sc Proof.} Consider first the case $\beta \geq 1$ and use
induction by $\beta$.
If $\beta=1$, (\ref{eq:lemma_statement}) is true because of
(\ref{eq:fd_derivatives}).

Assume (\ref{eq:lemma_statement}) is true for some $\beta = l \geq 1$.
Then, using
(\ref{eq:step_size}), (\ref{eq:fd_derivatives}) and
(\ref{eq:zero_derivative}), we get by direct calculation
(see Appendix~\ref{appendix:1} for details):
\begin{eqnarray}
  u_{\alpha+l+1}^{(J)}
=  u_{\alpha}^{(J)}
  + (u_{\alpha+l+1}^{(J)}-u_{\alpha+l}^{(J)})
  + (u_{\alpha+l  }^{(J)}-u_{\alpha  }^{(J)})
=
  u_{\alpha}^{(J)} +
  \sum_{i=1}^{p-J}
  \frac{\varepsilon^{i}}{i!}
  h_{\alpha+l+1,\alpha}^i u_{\alpha}^{(J+i)}
  + o(\varepsilon^{p-J}),                    \label{eq:lemma_1}
\end{eqnarray}
which proves the validity of (\ref{eq:lemma_statement})
for $\beta=l+1$.
Therefore, by induction, the Lemma is true for $\beta \geq 1$.

Consider now the case $\beta \leq -1$. We again perform
induction by $\beta$ starting from $\beta=-1$ and going
in the direction of decreasing $\beta$. To prove the
statements on each of the steps of this induction,
induction by $J$ will be used.

To prove (\ref{eq:lemma_statement}) for $\beta=-1$, consider
induction by $J$ from $J=p$ to $J=0$. First, for $J=p$
(\ref{eq:fd_derivatives}) implies
$u_{\alpha-1}^{(p)}=u_{\alpha}^{(p)}+o(1)$, which
means that  (\ref{eq:lemma_statement}) is valid when $\beta=-1$
and $J=p$. Assume that (\ref{eq:lemma_statement}) is true for
 $\beta=-1$ and $J=J_0+1,...,p$. Prove it for $J=J_0$.
Using (\ref{eq:step_size}), (\ref{eq:fd_derivatives}),
(\ref{eq:zero_derivative}) and induction hypothesis one
obtains by direct calculation (see Appendix~\ref{appendix:1} for details):
\begin{eqnarray}
  u_{\alpha-1}^{(J_0)} &=& u_{\alpha}^{(J_0)}
 -\sum_{k=1}^{p-J_0} \frac{\varepsilon^k h_{\alpha-1}^k}{k!}
  u_{\alpha-1}^{(J_0+k)} + o(\varepsilon^{p-J_0}) \nonumber \\
&=& u_{\alpha}^{(J_0)}
 -\sum_{k=1}^{p-J_0} \frac{\varepsilon^k h_{\alpha-1}^k}{k!}
  \left[
        \sum_{m=0}^{p-J_0} \frac{\varepsilon^m (-h_{\alpha-1})^m}{m!}
        u_{\alpha}^{(J_0+k+m)}
  \right]
 + o(\varepsilon^{p-J_0}) \label{eq:lemma_2} \\
&=&
  u_{\alpha}^{(J_0)}
 +\sum_{i=k}^{p-J_0}
  \varepsilon^{i}
  \frac{(-h_{\alpha-1})^{i}}{i!}
  u_{\alpha}^{(J_0+i)}
 + o(\varepsilon^{p-J_0}),    \nonumber
\end{eqnarray}
and therefore (\ref{eq:lemma_statement}) is true for $\beta=-1$
and all possible values of $J$.

Now, assume (\ref{eq:lemma_statement}) is true for all
 $\beta \geq \beta_0$, where $\beta_0 \leq -1$.
Prove it for $\beta=\beta_0-1$.
Again use induction by $J$. For $J=p$ (\ref{eq:lemma_statement})
immediately follows from (\ref{eq:fd_derivatives}). Suppose
(\ref{eq:lemma_statement}) is valid for $J=J_0+1,...,p$. Using
(\ref{eq:step_size}), (\ref{eq:fd_derivatives}),
(\ref{eq:zero_derivative}) and induction hypothesis prove
that it is true for $J=J_0$ (see Appendix~\ref{appendix:1}) :
\begin{eqnarray}
  u_{\alpha+\beta_0-1}^{J_0}
&=& \sum_{k=0}^{p-J_0}
  \frac{\varepsilon^{k} (-h_{\alpha+\beta_0-1})^k}{k!}
  u_{\alpha+\beta_0}^{(J_0+k)}
  + o(\varepsilon^{p-J_0})  \nonumber \\
&=& \sum_{k=0}^{p-J_0}
  \frac{\varepsilon^{k} (-h_{\alpha+\beta_0-1})^k}{k!}
  \left[
    \sum_{m=0}^{p-J_0-k}
    \frac{\varepsilon^m h_{\alpha+\beta_0,\alpha}^m}{m!}
    u_{\alpha}^{(J_0+k+m)}
  \right]
  + o(\varepsilon^{p-J_0}) \label{eq:lemma_3} \\
&=& u_{\alpha}^{(J_0)}
 +\sum_{i=1}^{p-J_0}
  \frac{\varepsilon^{i}}{i!}
  u_{\alpha}^{(J_0+i)}
  h_{\alpha+\beta_0-1,\alpha}^i
+ o(\varepsilon^{p-J_0}),  \nonumber
\end{eqnarray}
which proves (\ref{eq:lemma_statement}) for all possible $\beta \leq -1$
and $J$.

\noindent $\Box$
\mbox{ }\\
\\

\begin{lemma} \label{lemma:expansion_of_a_function}
  If $1 \leq \alpha \leq N$, and $\xi(x,u)$
  is a $p$ times continuously differentiable function,
  then, by virtue of (\ref{eq:step_size}), (\ref{eq:fd_derivatives}) and
  (\ref{eq:zero_derivative})
  \begin{eqnarray}
     \xi(x_{\alpha+1},u_{\alpha+1})
   - \xi(x_{\alpha  },u_{\alpha  })
   = \sum_{m=1}^{p} \frac{(\varepsilon h_\alpha)^m}{m!}
     D^m\xi(x_{\alpha  },u_{\alpha  }) + o(\varepsilon^p),
  \end{eqnarray}
  where  $D = \partial_x + u^{(1)} \partial_u + u^{(2)} \partial_{u^{(1)}} +
...$.
\end{lemma}
{\sc Proof.}
Using smoothness of the function $\xi(x, u)$
and rearranging terms in a Taylor expansion
(see Appendix~\ref{appendix:2}) write
\begin{eqnarray}
   \xi(x_{\alpha+1},u_{\alpha+1})
 = \sum_{m=0}^{p} \frac{(\varepsilon h_\alpha)^m}{m!}
   \left[
          \sum_{i=0}^{m}
          C_m^i \left(
                       \sum_{n=0}^{m-i}
                       \frac{(m-i)!}{n!} L^{(m-i)}_{(n)} \partial_u^n
                \right)
                \partial_x^i \xi(x_{\alpha},u_{\alpha})
   \right]
   + o(\varepsilon^p), \label{eq:lemma2_1}
\end{eqnarray}
where
\begin{eqnarray*}
  L^{(l)}_{(n)} = \sum_{l_1+...+l_n = l} L_{l_1}...L_{l_n}, \;\;\;\; n \neq 0,
\end{eqnarray*}
$L_l=u^{(l)}/l!$ and $L^{(l)}_{(0)}=\delta_{l,0}$.
To prove the Lemma, show that the expression in square brackets in
(\ref{eq:lemma2_1}) is equal to
\begin{eqnarray}
   D^m\xi(x_{\alpha  },u_{\alpha  })
=  \left[
        \partial_x + \sum_J u^{(J+1)} \partial_{u^{(J)}}
   \right]^m\xi(x_{\alpha  },u_{\alpha  })
=  \sum_{i=0}^{m} C^i_m \left(
                               \left[
                                      \sum_J u^{(J+1)} \partial_{u^{(J)}}
                               \right]^{m-i}
                        \right) \partial_x^i
   \xi(x_{\alpha  },u_{\alpha  }) \label{eq:lemma2_2}
\end{eqnarray}
Comparing equations (\ref{eq:lemma2_2}) and
(\ref{eq:lemma2_1}), one can see that it is sufficient to prove
that the expressions in round brackets in these equations
are equal for all values of $l = m-i$ varying between $0$
and $m$, ie
\begin{eqnarray}
    \sum_{n=0}^{l}
    \frac{l!}{n!} L^{(l)}_{(n)} \partial_u^n
    \xi(x_{\alpha  },u_{\alpha  })
 =  \left(
      \sum_J u^{(J+1)} \partial_{u^{(J)}}
    \right)^l
    \xi(x_{\alpha  },u_{\alpha  }) \label{eq:lemma2_3}
\end{eqnarray}
This equality is proven using induction by $l$. When
$l=0$, the equality becomes an identity. Assume that the
equality is true for $l=k < m$. By applying the
differential operator $\sum_J u^{(J+1)} \partial_{u^{(J)}}$
to both sides of (\ref{eq:lemma2_3}) written for
$l=k$ and manipulating the terms in the left--hand
side of the resulting equation (see Appendix~\ref{appendix:2}),
one obtains equation (\ref{eq:lemma2_3}) with
$l=k+1$. This completes the proof of the Lemma.

\noindent $\Box$
\mbox{ }\\
\\


Consider one of the equations (\ref{eq:FDE1}) corresponding
to $\alpha = \alpha_0$. Using equations (\ref{eq:step_size}),
(\ref{eq:fd_derivatives}) and Lemma~\ref{lemma:expansion}
equation (\ref{eq:FDE1}) may be rewritten as follows:
\begin{eqnarray}
& & L_{\alpha_0}(x_1,...,x_N,u_1,...,u_N) \nonumber\\
&=& L_{\alpha_0}
 \left(x_{\alpha_0} - \varepsilon \sum_{i=1}^{\alpha_0-1} h_i,\;
       ...,\;
       x_{\alpha_0},\;
       ...,\;
       x_{\alpha_0} + \varepsilon \sum_{i=\alpha_0}^{N-1} h_i,
 \right. \nonumber \\
& &
 \left.
       u_{\alpha_0} + \sum_{k=1}^p \frac{\varepsilon^k}{k!}
                       (-\sum_{i=1}^{\alpha_0-1} h_i)^k u_{\alpha_0}^{(k)},
       \;
       ...,\;
       u_{\alpha_0},\;
       ...,\;
       u_{\alpha_0} + \sum_{k=1}^p \frac{\varepsilon^k}{k!}
                       (\sum_{i=\alpha_0}^{N-1} h_i)^k u_{\alpha_0}^{(k)}
 \right)
+o(\varepsilon^p)
                               \label{eq:substitution}        \\
&=&L^{(0)}_{\alpha_0}(x_{\alpha_0},u_{\alpha_0}^{(0)},
                                 u_{\alpha_0}^{(1)},
                                 ...,
                                 u_{\alpha_0}^{(p)},
                                 h_1,...,h_{N-1})
+\sum_{i=1}^r \varepsilon^i
 L^{(i)}_{\alpha_0}(x_{\alpha_0},u_{\alpha_0}^{(0)},
                                 u_{\alpha_0}^{(1)},
                                 ...,
                                 u_{\alpha_0}^{(p)},
                                 h_1,...,h_{N-1})
+o(\varepsilon^r),  \nonumber
\end{eqnarray}
where $1 \leq r \leq p$ is some integer number.
\begin{definition}
 The system of algebraic equations (\ref{eq:FDE1}),
 (\ref{eq:step_size}), (\ref{eq:fd_derivatives}) and
 (\ref{eq:zero_derivative})
 is said to approximate
 equation (\ref{eq:ODE1}) at the node $x_{\alpha_0}$, if,
 for all values of $x_{\alpha_0}, u_{\alpha_0}^{(0)},
                    u_{\alpha_0}^{(1)}, ..., u_{\alpha_0}^{(p)},
                    h_1, ..., h_{N-1}$, we get
 \begin{eqnarray*}
   L^{(0)}_{\alpha_0}(x_{\alpha_0},u_{\alpha_0}^{(0)},
                                   u_{\alpha_0}^{(1)},
                                   ...,
                                   u_{\alpha_0}^{(p)},
                                   h_1,...,h_{N-1})
 = F(x_{\alpha_0}, u_{\alpha_0}^{(0)}, u_{\alpha_0}^{(1)}, ...,
     u_{\alpha_0}^{(n)}),
 \end{eqnarray*}
 where $F$ is the function from (\ref{eq:ODE1}).
\end{definition}

\begin{definition}
  A differential equation
  \begin{eqnarray}
    F(x, y, y^{(1)}, ..., y^{(n)})
    +\sum_{i=1}^r \varepsilon^i
      L^{(i)}_{\alpha_0}(x, y, y^{(1)}, ..., y^{(p)}, h_1,...,h_{N-1}) =0,
                                   \label{eq:differential_approximation}
  \end{eqnarray}
  where $1 \leq r \leq p$, $\varepsilon$ is a small parameter and
  $h_1,...,h_{N-1}$ are parameters (not small)
  is called a differential approximation of $r$-th order at the
  node $x_{\alpha_0}$ for the system of equations (\ref{eq:FDE1}),
  (\ref{eq:step_size}), (\ref{eq:fd_derivatives}) and
  (\ref{eq:zero_derivative}). If $r=1$, the differential equation
  (\ref{eq:differential_approximation}) is called the first
  differential approximation at the node $x_{\alpha_0}$.
\end{definition}

Assuming that the set of equations (\ref{eq:FDE1}),
 (\ref{eq:step_size}), (\ref{eq:fd_derivatives}) and
 (\ref{eq:zero_derivative}) approximates the differential
 equation (\ref{eq:ODE1}), one can rewrite the
 former set of algebraic equations as follows:
\begin{eqnarray}
  F(x_{\alpha^\prime},u_{\alpha^\prime}^{(0)},
                                 u_{\alpha^\prime}^{(1)},
                                 ...,
                                 u_{\alpha^\prime}^{(n)})
+\sum_{i=1}^r \varepsilon^i
 L^{(i)}_{\alpha^\prime}(x_{\alpha^\prime},u_{\alpha^\prime}^{(0)},
                                 u_{\alpha^\prime}^{(1)},
                                 ...,
                                 u_{\alpha^\prime}^{(p)},
                                 h_1,...,h_{N-1})
&=&o(\varepsilon^r),
                           \label{eq:FDE2}\\
  x_{\alpha+1}-x_{\alpha} - \varepsilon h_\alpha &=& o(\varepsilon^r),
                           \label{eq:step_size2} \\
  u_{\alpha+1}^{(J)}-u_{\alpha}^{(J)} -
  \sum_{k=1}^{r} \frac{\varepsilon^k}{k!} h_\alpha^k u_{\alpha}^{(J+k)}
  &=& o(\varepsilon^r),      \label{eq:fd_derivatives2} \\
  u_{\alpha^\prime}^{(0)} - u_{\alpha^\prime} &=& 0,
                           \label{eq:zero_derivative2}
\end{eqnarray}
where $\alpha = 1,...,N-1, \;\;\;\; \alpha^\prime = 1,...,N, \;\;\;\;
1 \leq r \leq p, \;\;\;\; p \geq n, \;\;\;\; J = 0,...,p$.

As a result of the above, we obtain the set of
equations (\ref{eq:FDE2}), (\ref{eq:step_size2}),
(\ref{eq:fd_derivatives2}), (\ref{eq:zero_derivative2}) with a small
parameter $\varepsilon$. In the following, approximate symmetries
of this system and their relation to the symmetries of the
differential equation (\ref{eq:ODE1}) will be studied.
\mbox{ }\\
\\
{\bf Example.} Consider a general  first order differential equation
resolved with respect to the first derivative:
\begin{eqnarray}
  F(x,y,y^{(1)}) \equiv y^{(1)}-f(x, y) = 0,\;\;\;\; x \in (0, 1)
                                        \label{eq:ODE_EXAMPLE1}
\end{eqnarray}
and its finite--difference approximation based upon forward
differencing:
\begin{eqnarray}
  L_\alpha(x_\alpha,x_{\alpha+1},u_\alpha,u_{\alpha+1})
\equiv
\frac{u_{\alpha+1}-u_\alpha}{x_{\alpha+1}-x_\alpha}-f(x_\alpha, u_\alpha)
=
0,
\;\;\;\;\alpha = 1,...,N-1.              \label{eq:APPROX_EXAMPLE1}
\end{eqnarray}
Set $p=2$ and obtain the system (\ref{eq:FDE2}), (\ref{eq:step_size2}),
(\ref{eq:fd_derivatives2}), (\ref{eq:zero_derivative2}) corresponding
to the finite--difference equation (\ref{eq:APPROX_EXAMPLE1}). Using
(\ref{eq:step_size2}) and (\ref{eq:fd_derivatives2}) in
(\ref{eq:APPROX_EXAMPLE1}), we get:
\begin{eqnarray*}
  u_{\alpha}^{(1)}
- f(x_{\alpha}, u_{\alpha}^{(0)})
+ \frac{\varepsilon h_{\alpha}}{2} u_{\alpha}^{(2)}
&=&o(\varepsilon), \\
  x_{\alpha+1}-x_{\alpha} - \varepsilon h_\alpha &=& o(\varepsilon), \\
  u_{\alpha+1}^{(J)}-u_{\alpha}^{(J)} -
  \varepsilon h_\alpha u_{\alpha}^{(J+1)}
  &=& o(\varepsilon), \\
  u_{\alpha^\prime}^{(0)} - u_{\alpha^\prime} &=& 0,
\end{eqnarray*}
where $\alpha = 1,...,N-1, \;\;\;\; \alpha^\prime = 1,...,N, \;\;\;\;
J = 0,1,2$.

According to the definitions given, the above system
of finite--difference equations approximates
a differential equation (\ref{eq:ODE_EXAMPLE1}) at all nodes
 $x_\alpha$, $\alpha = 1,...,N-1$. At any of these nodes
the first differential
approximation corresponding to (\ref{eq:APPROX_EXAMPLE1}) is
\begin{eqnarray*}
   y^{(1)} - f(x, y) + \frac{\varepsilon h_{\alpha}}{2} y^{(2)}  = 0.
\end{eqnarray*}
\mbox{ } \\
\\
As can be seen from this Example, the term with a small
parameter, appearing in the first differential approximation,
is linear with respect to $h_\alpha$.
The following Theorem gives the necessary and sufficient condition
that equation (\ref{eq:differential_approximation}) with $r=1$
needs to satisfy in order to be a first differential approximation
of some finite--difference approximation of equation
(\ref{eq:ODE1}).
\begin{theorem} \label{theorem:3}
Let $N >  p > n$, then the equation
\begin{eqnarray}
    F(x, y, y^{(1)}, ..., y^{(n)})
    + \varepsilon
      L^{(1)}_{\alpha_0}(x, y, y^{(1)}, ..., y^{(p)}, h_1,...,h_{N-1}) =0,
                                  \label{eq:1st_differential_approximation}
\end{eqnarray}
is a first differential approximation of some finite--difference
approximation of equation (\ref{eq:ODE1}) at the node $x_{\alpha_0}$
if and only if
\begin{eqnarray}
  X( \varepsilon L^{(1)}_{\alpha_0} ) \equiv 0,  \label{eq:theorem3_main}
\end{eqnarray}
where
\begin{eqnarray}
  X = - \varepsilon \partial_\varepsilon
      + \sum_{i=1}^n h_{\alpha_i} \partial_{\alpha_i}. \label{eq:theorem3_X}
\end{eqnarray}
\end{theorem}
{\sc Proof.} 1) Let the condition (\ref{eq:theorem3_main}) be satisfied.
The following describes the procedure resulting in a finite--difference
approximation which has
(\ref{eq:1st_differential_approximation}) as its first differential
approximation.

The differential operator $X$ in (\ref{eq:theorem3_X}) can be treated
as an infinitesimal operator of the o.p.g. acting in the space
of $p+N+2$ variables
$\varepsilon, x, y, y^{(1)}, ..., y^{(p)}, h_1,..., h_{N-1}$.
As one can easily see,
 $x, y, y^{(1)}, ..., y^{(p)},$$ \varepsilon h_1,..., \varepsilon h_{N-1}$
are $p+N+1$ functionally independent invariants of this o.p.g., and
 $\varepsilon L^{(1)}_{\alpha_0}$ is also an invariant of the same
group because of (\ref{eq:theorem3_main}). Using Theorem 2.17
(\cite{OLVER}, p.88) and formally substituting
 $x_{\alpha + 1} - x_{\alpha}$ for each expression
 $\varepsilon h_{\alpha}$ and $x_{\alpha_0}$ for $x$,
we establish that
\begin{eqnarray*}
   \varepsilon L^{(1)}_{\alpha_0}
 &=& \tilde{f}_1(x, y, y^{(1)}, ..., y^{(p)},
               \varepsilon h_1,..., \varepsilon h_{N-1})
 = \tilde{f}_1(x_{\alpha_0}, y, y^{(1)}, ..., y^{(p)},
               x_2 - x_1, ..., x_N - x_{N-1})  \\
&=&\bar{f}_1(y, y^{(1)}, ..., y^{(p)},
               x_1, ..., x_N)
\end{eqnarray*}
for some smooth functions $\tilde{f}$ and $\bar{f}$.

Utilizing (\ref{eq:lemma_statement}) and the fact that $N > p > n$
substitute $y, y^{(1)}, ..., y^{(p)}$ with their respective
finite-difference approximations of the first order of accuracy
on the grid composed of the nodes $x_1, ...,x_N$ with $u_\alpha$
denoting the value of the grid function at the node $x_\alpha$:
\begin{eqnarray*}
  \bar{f}_1(y, y^{(1)}, ..., y^{(p)},x_1, ..., x_N)
= \hat{f}_1(u_1, ..., u_N,x_1, ..., x_N) + o(1)
\end{eqnarray*}
for some function $\hat{f}_1$.
Acting in a similar way, substitute the arguments
 $y, y^{(1)}, ..., y^{(n)}$ with their finite--difference
approximations of the order higher than the first in the
expression for $F(x,y, y^{(1)}, ..., y^{(n)})$,
and also substitute $x$ with $x_{\alpha_0}$ in the same
expression:
\begin{eqnarray*}
   F(x,y, y^{(1)}, ..., y^{(n)})
= \hat{f}_0(u_1, ..., u_N,x_1, ..., x_N) + o(\varepsilon)
\end{eqnarray*}
for some function $\hat{f}_0$.

Now chose the finite--difference approximation at the node
$x_{\alpha_0}$ to be
\begin{eqnarray}
       L(u_1, ..., u_N,x_1, ..., x_N)
\equiv \hat{f}_0(u_1, ..., u_N,x_1, ..., x_N)
     + \hat{f}_1(u_1, ..., u_N,x_1, ..., x_N)=0,
                         \label{eq:required_approximation}
\end{eqnarray}
where $\hat{f}_0$ and $\hat{f}_1$ are the previously constructed
functions.

Reversing the above speculation, we can see that, by virtue
of (\ref{eq:lemma_statement}) and (\ref{eq:step_size}):
\begin{eqnarray*}
  L(u_1, ..., u_N,x_1, ..., x_N)
= F(x_{\alpha_0},u_{\alpha_0}, u^{(1)}_{\alpha_0}, ..., u^{(n)}_{\alpha_0})
+ \varepsilon
  L^{(1)}_{\alpha_0}(x_{\alpha_0},u_{\alpha_0}, u^{(1)}_{\alpha_0}, ...,
                     u^{(n)}_{\alpha_0}, h_1,...,h_{N-1})
+ o(\varepsilon),
\end{eqnarray*}
that is,
(\ref{eq:1st_differential_approximation})
is a first differential approximation of (\ref{eq:required_approximation}).

2) Let (\ref{eq:FDE1}) with $\alpha = \alpha_0$ approximate
(\ref{eq:ODE1}). Introduce the following notation:
\begin{eqnarray*}
& &\tilde{L}_{\alpha_0}(x_{\alpha_0},u_{\alpha_0}^{(0)},
                                 u_{\alpha_0}^{(1)},
                                 ...,
                                 u_{\alpha_0}^{(p)},
                                 \varepsilon h_1,...,\varepsilon h_{N-1}) \\
&=&L_{\alpha_0}
 \left(x_{\alpha_0} - \varepsilon \sum_{i=1}^{\alpha_0-1} h_i,\;
       ...,\;
       x_{\alpha_0},\;
       ...,\;
       x_{\alpha_0} + \varepsilon \sum_{i=\alpha_0}^{N-1} h_i,
 \right.  \\
& &
 \left.
       u_{\alpha_0} + \sum_{k=1}^p \frac{\varepsilon^k}{k!}
                       (-\sum_{i=1}^{\alpha_0-1} h_i)^k u_{\alpha_0}^{(k)},
       \;
       ...,\;
       u_{\alpha_0},\;
       ...,\;
       u_{\alpha_0} + \sum_{k=1}^p \frac{\varepsilon^k}{k!}
                       (\sum_{i=\alpha_0}^{N-1} h_i)^k u_{\alpha_0}^{(k)}
 \right).
\end{eqnarray*}
Notice that it follows from this definition and (\ref{eq:substitution}),
that
$X \tilde{L}_{\alpha_0} \equiv 0$ and also that
$L^{(1)}_{\alpha_0} = (\partial_\varepsilon \tilde{L}_{\alpha_0})|_{\varepsilon
= 0}$.

Using these relations, write
\begin{eqnarray*}
  \sum_{i=1}^n h_{\alpha_i} \partial_{\alpha_i}
    (\partial_\varepsilon \tilde{L}_{\alpha_0})
= \partial_\varepsilon
  (\sum_{i=1}^n h_{\alpha_i} \partial_{\alpha_i} \tilde{L}_{\alpha_0} )
= \partial_\varepsilon(\varepsilon \partial_\varepsilon \tilde{L}_{\alpha_0})
= \partial_\varepsilon \tilde{L}_{\alpha_0}
+ \varepsilon \partial_\varepsilon^2 \tilde{L}_{\alpha_0},
\end{eqnarray*}
Taking the limit $\varepsilon \rightarrow 0$ in the latter equation
results in (\ref{eq:theorem3_X}).

\noindent $\Box$
\mbox{ }\\
\\

\subsection{Approximate groups leaving finite--difference approximation
            invariant}

Consider a one--parameter group of transformations acting on
the variables $x$ and $y$ and represented by its infinitesimal
operator $X = \xi(x,y) \partial_x + \varphi(x,y) \partial_y$.
Action of this group  is extended to the derivatives
$y^{(1)}, ..., y^{(n)}$ using prolongation formulae \cite{OLVER}:
\begin{eqnarray}
  X = \xi(x,y) \partial_x + \varphi(x,y) \partial_y
    + \sum_{i=1}^n \varphi^{(i)} \partial_{y^{(i)}},
         \label{eq:original_group}
\end{eqnarray}
where
\begin{eqnarray}
  \varphi^{(i)} &=& D\varphi^{(i-1)} - y^{(i)} D \xi,
                                   \label{eq:prolongation} \\
  D &=& \partial_x + y^{(1)} \partial_y + y^{(2)} \partial_{y^{(1)}} + ...,
                                   \nonumber               \\
 \varphi^{(0)} &=&  \varphi.       \nonumber
\end{eqnarray}

Action of this group of transformations can also be naturally
extended to the ``finite--difference'' variables
 $x_\alpha$ and $u_\alpha$, ($\alpha = 1,...,N$) as
\begin{eqnarray}
   X^{(\Delta)} = \sum_{\alpha = 1}^N
         \xi(x_\alpha,u_\alpha) \partial_{x_\alpha}
   + \varphi(x_\alpha,u_\alpha) \partial_{u_\alpha}.
         \label{eq:naturally_extended_group}
\end{eqnarray}
In order to extend action of this group to the
variables $u^{(J)}_1, ..., u^{(J)}_{N}$, ($J=1, ..., 2p$) and
 $h_1, ..., h_{N-1}$ while preserving ``locality'' of the prolongation
 formulae, consider an a.o.p.g. represented by its infinitesimal
 operator
\begin{eqnarray}
 X^{(\Delta)} &=& \sum_{k=0}^{r} \varepsilon^k {\mX_k}\mbox{}^{(\Delta)}
        + o(\varepsilon^r),
                          \label{eq:approximate_group}\\
 {\mX_k}\mbox{}^{(\Delta)} &=& \sum_{\alpha = 1}^N
         \left[
                   \mxi_k(x_\alpha,u_\alpha) \partial_{x_\alpha}
             + \mvarphi_k(x_\alpha,u_\alpha) \partial_{u_\alpha}
             + \sum_{i=1}^{2r}
               {\mvarphi_k}\mbox{}^{(i)}_\alpha \partial_{u_\alpha^{(i)}}
        \right]
       +\sum_{\alpha = 1}^{N-1}{\msigma_k}\mbox{}_\alpha \partial_{h_\alpha}.
        \nonumber
\end{eqnarray}

In the following, we will be interested in those of the groups
(\ref{eq:approximate_group}), which leave invariant the set of
approximating finite--difference equations (\ref{eq:FDE2}),
(\ref{eq:step_size2}) and (\ref{eq:fd_derivatives2}).
The following Theorem establishes the analogues of the prolongation
formulae (\ref{eq:prolongation}) for the finite--difference
equations:
\begin{theorem} \label{theorem:fd_prolongation}
  Equations (\ref{eq:step_size2}) and (\ref{eq:fd_derivatives2})
  are invariant with respect to the a.o.p.g.
  (\ref{eq:approximate_group}) if and only if
  \begin{eqnarray}
    \msigma_k\mbox{}_\alpha &=& \sum_{l=1}^{k+1}
                                \frac{h_\alpha^l}{l!}
                                D^l \mxi_{k+1-l}(x_\alpha,u_\alpha),
                                \label{eq:stepsize_prolongation} \\
    {\mvarphi_k}\mbox{}^{(J+1)}_\alpha &=&
                                D {\mvarphi_k}\mbox{}^{(J)}_\alpha
                              - u_\alpha^{(J+1)}
                                D \mxi_{k}(x_\alpha,u_\alpha),
                                \label{eq:fd_prolongation}
  \end{eqnarray}
  where $0 \leq k \leq r-1$.
\end{theorem}
{\sc Proof.}
Using the infinitesimal criterion of approximate invariance
(Theorem~\ref{theorem:infinitesimal_criterion}), apply
the operator (\ref{eq:approximate_group}) to equation
(\ref{eq:step_size2}) to obtain:
\begin{eqnarray*}
  \mxi_0\mbox{}_{\alpha+1} - \mxi_0\mbox{}_{\alpha} +
  \sum_{k=1}^r
    \varepsilon^k (\mxi_k\mbox{}_{\alpha+1} - \mxi_k\mbox{}_{\alpha}
                 - \msigma_{k-1}\mbox{}_\alpha) =
    o(\varepsilon^r).
\end{eqnarray*}
Using Lemma~\ref{lemma:expansion_of_a_function}, rewrite the left--hand
side of the above equation as follows:
\begin{eqnarray*}
  \sum_{k=0}^r \varepsilon^k (\mxi_k\mbox{}_{\alpha+1}
                            - \mxi_k\mbox{}_{\alpha} )
 -\sum_{k=1}^r \varepsilon^k \msigma_{k-1}\mbox{}_\alpha
= \sum_{k=0}^r  \sum_{l=1}^r \varepsilon^{k+l}
    \frac{h_\alpha^l}{l!} D^l \mxi_k\mbox{}_{\alpha}
 -\sum_{k=1}^r \varepsilon^k \msigma_{k-1}\mbox{}_\alpha
 + o(\varepsilon^r)\\
= \sum_{i=1}^r \varepsilon^i
  \left\{
    \sum_{l=1}^i \frac{h_\alpha^l}{l!} D^l \mxi_{i-l}\mbox{}_{\alpha}
    - \msigma_{i-1}\mbox{}_\alpha
  \right\}
+ o(\varepsilon^r),
\end{eqnarray*}
which immediately results in (\ref{eq:stepsize_prolongation}).

Similarly, having applied Theorem~\ref{theorem:infinitesimal_criterion}
to equation (\ref{eq:fd_derivatives2}), one obtains
\begin{eqnarray*}
 \sum_{k=0}^r \varepsilon^k
              \left\{
                 \mvarphi_k\mbox{}^{(J)}_{\alpha+1} -
\mvarphi_k\mbox{}^{(J)}_{\alpha}
                 -\sum_{i=1}^r \varepsilon^i
                 \left[
                     \frac{h_\alpha^i}{i!} \mvarphi_k\mbox{}_\alpha^{(J+i)}
                    +\frac{h_\alpha^{i-1}}{(i-1)!} \msigma_k\mbox{}_\alpha
u_\alpha^{(J+i)}
                 \right]
              \right\} = o(\varepsilon^r),
\end{eqnarray*}
which must hold by virtue of (\ref{eq:step_size2}) and
(\ref{eq:fd_derivatives2}).
Applying Lemma~\ref{lemma:expansion_of_a_function} and
(\ref{eq:stepsize_prolongation}),
transform the left--hand side of the above equation as follows:
\begin{eqnarray*}
& & \sum_{i=1}^r \sum_{k=0}^r \varepsilon^{i+k}
    \left\{
           \frac{h_\alpha^i}{i!} D^i \mvarphi_k\mbox{}_\alpha^{(J)}
          -\frac{h_\alpha^i}{i!} \mvarphi_k\mbox{}_\alpha^{(J+i)}
          -\frac{h_\alpha^{i-1}}{(i-1)!} \msigma_k\mbox{}_\alpha
u_\alpha^{(J+i)}
    \right\} \\
&=& \sum_{l=1}^r \varepsilon^l \sum_{i=1}^l
    \left\{
           \frac{h_\alpha^i}{i!} D^i \mvarphi_{l-i}\mbox{}_\alpha^{(J)}
          -\frac{h_\alpha^i}{i!} \mvarphi_{l-i}\mbox{}_\alpha^{(J+i)}
          -\frac{h_\alpha^{i-1}}{(i-1)!} \msigma_{l-i}\mbox{}_\alpha
u_\alpha^{(J+i)}
    \right\} + o(\varepsilon^r)\\
&=& \sum_{l=1}^r \varepsilon^l \sum_{i=1}^l \frac{h_\alpha^i}{i!}
    \left\{
            D^i \mvarphi_{l-i}\mbox{}_\alpha^{(J)}
          - \mvarphi_{l-i}\mbox{}_\alpha^{(J+i)}
          - \sum_{n=1}^{i} C_i^{n-1} u_\alpha^{(J+n)}
                           D^{i-n+1} \mxi_{l-i}\mbox{}_\alpha
    \right\} + o(\varepsilon^r).
\end{eqnarray*}
Equating the obtained expression to $o(\varepsilon^r)$ results in
the following set of equalities:
\begin{eqnarray}
    \sum_{i=1}^l \frac{h_\alpha^i}{i!}
    \left\{
            D^i \mvarphi_{l-i}\mbox{}_\alpha^{(J)}
          - \mvarphi_{l-i}\mbox{}_\alpha^{(J+i)}
          - \sum_{n=1}^{i} C_i^{n-1} u_\alpha^{(J+n)}
                           D^{i-n+1} \mxi_{l-i}\mbox{}_\alpha
    \right\} = 0,  \label{eq:preliminary}
\end{eqnarray}
where $l=1,2,...,r$.

Use induction by $l$ to prove (\ref{eq:fd_prolongation}) for all $0 \leq k \leq
r-1$.
Setting $l=1$ in (\ref{eq:preliminary}) immediately results in
(\ref{eq:fd_prolongation}) with $k=0$. Assume (\ref{eq:fd_prolongation})
holds for all $k = 0,1,...,k_1$ and prove it for $k=k_1+1 \leq r-1$.
Consider (\ref{eq:preliminary}) with $l=k_1+2$ . It is easy to
see that the terms in the brackets, corresponding to $i=2,...,k_1+2$ nullify
because of the induction hypothesis, while the remaining term,
being equated to $0$, gives (\ref{eq:fd_prolongation}) with $k=k_1+1$.

\noindent $\Box$
\mbox{ }\\
\\

One can immediately notice the similarity between the prolongation
formulae for the finite--difference derivatives
(\ref{eq:fd_prolongation}) and their continuous analogues
(\ref{eq:prolongation}). This similarity allows one to analyze
and to establish some important group properties of the finite--difference
approximations in terms of the group properties of their
differential approximations.
Really, consider also the system of algebraic equations
with a small parameter
(\ref{eq:FDE2}),
(\ref{eq:step_size2}),(\ref{eq:fd_derivatives2}) and
(\ref{eq:zero_derivative2})  resulting from some discretization of the
equation (\ref{eq:ODE1}). According to Theorem~\ref{theorem:fd_prolongation},
the necessary and sufficient condition for the subsystem
(\ref{eq:step_size2}),(\ref{eq:fd_derivatives2})
to remain invariant under the action of the group (\ref{eq:approximate_group})
is that ${\displaystyle {\mvarphi_k}\mbox{}^{(J)}_\alpha}$ and
 ${\displaystyle\msigma_k\mbox{}_\alpha}$ are given by
(\ref{eq:stepsize_prolongation}) and (\ref{eq:stepsize_prolongation}).
Using these expressions and applying the operator (\ref{eq:approximate_group})
to equation (\ref{eq:FDE2}), one can see that the obtained
system of determining equations is the same as the one resulting
from the application of the infinitesimal criterion of invariance
to the differential approximation (\ref{eq:differential_approximation})
provided that action of the group is extended on $h_\alpha$
according to (\ref{eq:stepsize_prolongation}).

\mbox{ } \\
\\
{\bf Example.} Consider again an ordinary differential equation (\ref{eq:ODE1})
and the approximating set of algebraic equations
(\ref{eq:FDE2})--(\ref{eq:zero_derivative2}) with $r=1$. The
first differential approximations corresponding to the nodes $x_\alpha$,
$\alpha=1,...,N$
read:
\begin{eqnarray}
    F(x,y,y^{(1)},...,y^{(n)})
  + \varepsilon L_\alpha^{(1)}(x, y, y^{(1)},...,y^{(p)}, h_1,...,h_{N-1}) = 0.
  \label{eq:xxx}
\end{eqnarray}
Consider the following a.o.p.g.
\begin{eqnarray}
  X = \sum_{k=0}^1 \varepsilon^k
      \left(
            \mX_k + \sum_{\alpha=1}^{N-1} \msigma_k\mbox{}_\alpha
\partial_{h_\alpha}
      \right)
    + o(\varepsilon), \label{eq:xx}
\end{eqnarray}
where
\begin{eqnarray}
  \mX_k &=& \mxi_k(x,y) \partial_x + \mvarphi_k(x,y) \partial_y +
          \sum_{i=1}^{2p} \mvarphi_k\mbox{}^{(i)} \partial_{y^{(i)}}, \nonumber
\\
  \mvarphi_k\mbox{}^{(i)} &=& D \mvarphi_k\mbox{}^{(i-1)} - y^{(i)} D \mxi_k,
                                      \label{eq:x} \\
  \msigma_k\mbox{}_\alpha &=& \sum_{l=1}^{k+1} \frac{h_\alpha^l}{l!} D
\mxi_{k+1-l},
                                                                      \nonumber
\end{eqnarray}
that is, in addition to conventional prolongation formulae (\ref{eq:x}),
the relation (\ref{eq:stepsize_prolongation}) is used to extend the action of
the
group on the parameters $h_\alpha$.

Using Theorem~\ref{theorem:infinitesimal_criterion} and the methodology
outlined in the Example following this theorem, write the necessary
and sufficient conditions, under which equation (\ref{eq:xxx})
is invariant with respect to the a.o.p.g. (\ref{eq:xx}),
in the following way:
\begin{eqnarray}
  \mX_0 F - \lambda F &=& 0, \label{eq:xxxx} \\
  \left.
  \left(
    \mX_1 F
    + (\mX_0 + \sum_{\alpha=1}^{N-1} \msigma_k\mbox{}_\alpha
\partial_{h_\alpha})
      L_\alpha^{(1)}
    - \lambda L_\alpha^{(1)}
  \right)
  \right|_{F=0}  &=& 0,        \label{eq:xxxxx}
\end{eqnarray}
where $\lambda = \lambda(x,y,y^{(1)},...,y^{(n)})$ is some smooth
function and the fact that $F$ does not depend on $h_\alpha$ has
been taken into account. It follows from Theorem~\ref{theorem:3}
that $L_\alpha^{(1)}$ must be a linear function of the parameters
$h_\alpha$. Using this fact and also noticing that the first term
in (\ref{eq:xxxxx}) does not depend on $h_\alpha$ and therefore
should be equated to zero separately, rewrite (\ref{eq:xxxx})
and (\ref{eq:xxxxx}) as follows:
\begin{eqnarray}
  \mX_0 F - \lambda F &=& 0, \label{eq:o} \\
  \left.
  \left(
        \mX_0  L_\alpha^{(1)}
      + (D \mxi_0 - \lambda) L_\alpha^{(1)}
  \right)
  \right|_{F=0}   &=& 0,        \label{eq:oo} \\
    (\mX_1 F)|_{F=0}   &=& 0,        \label{eq:ooo}
\end{eqnarray}

The condition (\ref{eq:o}) implies that ${\displaystyle \mX_0}$
must be admitted exactly by the original differential equation
(\ref{eq:ODE1}). This condition also allows one to determine the
function $\lambda$. If the additional condition (\ref{eq:oo}) is
satisfied, the group ${\displaystyle \mX_0}$ can be corrected
by adding a small perturbation ${\displaystyle \varepsilon \mX_1}$
to it, so that the resulting corrected group leaves the first
differential equation (\ref{eq:xxx}) invariant. Finally, (\ref{eq:ooo})
shows that the correction ${\displaystyle \mX_1}$ may be any
group admitted by the original differential equation (\ref{eq:ODE1}),
or it can also be an operator with all zero components.
\mbox{ }\\
\\

\begin{definition}
A group (\ref{eq:original_group}), leaving the differential equation
(\ref{eq:ODE1}) invariant, is called stable with respect to the
approximation resulting in (\ref{eq:FDE2}),
(\ref{eq:step_size2}),(\ref{eq:fd_derivatives2}) and
(\ref{eq:zero_derivative2}) if and only if there exists an infinitesimal
operator ${\displaystyle \mX_1}$, such that the corresponding a.o.p.g.
${\displaystyle \mX_0\mbox{}^{(\Delta)}+\varepsilon
\mX_1\mbox{}^{(\Delta)}+o(\varepsilon)}$
leaves the system of equations (\ref{eq:FDE2}),
(\ref{eq:step_size2}),(\ref{eq:fd_derivatives2}) and
(\ref{eq:zero_derivative2}) invariant up to the $o(\varepsilon)$
terms, ie the a.o.p.g.
${\displaystyle \mX_0\mbox{}^{(\Delta)}+\varepsilon
\mX_1\mbox{}^{(\Delta)}+o(\varepsilon)}$ leaves
invariant the system  (\ref{eq:FDE2}),
(\ref{eq:step_size2}),(\ref{eq:fd_derivatives2}) and
(\ref{eq:zero_derivative2}), in which $r$ is set to be $1$.
\end{definition}

Using this notion, the results of this Section can be summarized
in the following Theorem.
\begin{theorem}
A group (\ref{eq:original_group}) leaving the differential equation
(\ref{eq:ODE1}) invariant is stable with respect to the
approximation which has the first differential approximation
(\ref{eq:1st_differential_approximation})
if and only if
  \begin{eqnarray}
    \left.
      \left(
           \mX_0 L_\alpha^{(1)} + (D(\mxi_0) - \lambda) L_\alpha^{(1)}
      \right)
    \right|_{F(x,y,y^{(1)},...,y^{(n)})=0}
    = 0
                                           \label{eq:stability_condition}
  \end{eqnarray}
for all $1 \leq \alpha \leq N$.
\end{theorem}

\subsection{Examples}
{\bf Example.} Consider the following ordinary differential equation:
\begin{eqnarray*}
   F(x, y, y^\prime) \equiv (y -x) y^\prime + x + y = 0, \\
   y = y (x), \;\;\;\; x \in (a, b)
\end{eqnarray*}
and the group
\begin{eqnarray*}
   X = y \partial_x - x \partial_y -
      (1-(y^\prime)^2) \partial_{y^\prime}
      - 3 y^\prime y^{\prime \prime} \partial_{y^{\prime \prime}}.
\end{eqnarray*}
Since $X F = -y^\prime F$, the above group leaves the equation invariant
and $\lambda = - y^\prime$.

Consider also the following finite--difference approximation of this
differential equation:
\begin{eqnarray*}
  (y_\alpha - x_\alpha) \frac{y_{\alpha+1}-y_\alpha}{x_{\alpha+1}-x_\alpha}
 + x_\alpha + y_\alpha = 0,
\end{eqnarray*}
where $x_\alpha \in (a, b)$ and $\alpha = 1, ..., N$.

The first differential approximation corresponding to this finite--difference
approximation reads:
\begin{eqnarray*}
  (y -x) y^\prime + x + y + \varepsilon \frac{h_\alpha}{2}
    y^{\prime \prime} (y-x) = o(\varepsilon),
\end{eqnarray*}
and therefore, $L_\alpha^{(1)} = \frac{h_\alpha}{2} y^{\prime \prime} (y-x)$.

{}From the above we get:
\begin{eqnarray*}
   X L^{(1)}_\alpha + (D(\xi) - \lambda) L^{(1)}_\alpha
= -\frac{h_\alpha}{2} y^{\prime \prime} ( (y -x) y^\prime + x + y) = 0,
\end{eqnarray*}
if $(y -x) y^\prime + x + y=0$.

Therefore the considered group is stable.
\mbox{ } \\
\\
{\bf Example.}The ordinary differential equation
\begin{eqnarray*}
   F(x, y, y^\prime) \equiv y^\prime -( x + 1 ) y^2 -y = 0, \\
   y = y (x), \;\;\;\; x \in (a, b)
\end{eqnarray*}
is invariant with respect to the group
\begin{eqnarray*}
   X = (y + x y^2) \partial_y
      +(y^\prime + y^2 + 2 x y y^\prime) \partial_{y^\prime}
      +(y^{\prime \prime} +4 y y^\prime +2 x (y^\prime)^2 + 2 x y y^{\prime
\prime})
                       \partial_{y^{\prime \prime}},
\end{eqnarray*}
because $X F = (1+2xy) F$, and therefore,  $\lambda = 1+2xy$.

Consider also the following finite--difference approximation of this
differential equation:
\begin{eqnarray*}
   \frac{y_{\alpha+1}-y_\alpha}{x_{\alpha+1}-x_\alpha}
 - (x_\alpha + 1) y_\alpha^2 - y_\alpha = 0,
\end{eqnarray*}
where $x_\alpha \in (a, b)$ and $\alpha = 1, ..., N$.

Applying the procedure described above, one obtains the following
first differential approximation corresponding to this finite--difference
approximation:
\begin{eqnarray*}
  y^\prime -( x + 1 ) y^2 -y + \varepsilon \frac{h_\alpha}{2} y^{\prime \prime}
=
  o(\varepsilon),
\end{eqnarray*}
and $L_\alpha^{(1)} = \frac{h_\alpha}{2} y^{\prime \prime} (y-x)$.

{}From the above we get:
\begin{eqnarray*}
   X L^{(1)}_\alpha + (D(\xi) - \lambda) L^{(1)}_\alpha
=  h_\alpha^2 (1+(x+1)y) (2+x+xy(x+1)),
\end{eqnarray*}
and the expression on the right--hand side is not
identically zero when $y^\prime -( x + 1 ) y^2 -y = 0$.

Therefore the considered group is not stable.

\mbox{ } \\
\\
{\bf Example.} Build a finite--difference approximation for the
differential equation in the previous example, such that the group
considered in the same example becomes stable.

First, find a first differential approximation of such a finite--difference
approximation:
\begin{eqnarray*}
  y^\prime -( x + 1 ) y^2 -y + \varepsilon
L_\alpha^{(1)}(x,y,y^\prime,y^{\prime \prime},h_\alpha)
  = o(\varepsilon)
\end{eqnarray*}

 $L_\alpha^{(1)}$ can be sought as
\begin{eqnarray*}
  L_\alpha^{(1)}(x,y,y^\prime,y^{\prime \prime},h_\alpha)
= \frac{h_\alpha}{2} y^{\prime \prime} + h_\alpha L^{*}(x,y).
\end{eqnarray*}

The stability condition (\ref{eq:stability_condition}) then results
in the following equation for $L^{*}(x,y)$:
\begin{eqnarray*}
  (1+xy)^2 \left(\frac{L^{*}}{y(1+xy)}\right)_y + (1+(x+1)y) (2+x+xy(x+1)) = 0,
\end{eqnarray*}
which gives
\begin{eqnarray*}
  L^{*} = - \frac{(x+1)^2}{x} y^2 (1+xy) - \frac{y}{x^2} + C(x) y (1+xy),
\end{eqnarray*}
where $C(x)$ is an arbitrary function.

Therefore, the first differential approximation of the stable
finite--difference
approximation must read:
\begin{eqnarray*}
  y^\prime -( x + 1 ) y^2 - y
+ \varepsilon h_\alpha
  \left(
         \frac{y^{\prime \prime}}{2}
       - \frac{(x+1)^2}{x} y^2 (1+xy) - \frac{y}{x^2} + C(x) y (1+xy)
  \right)
= o(\varepsilon).
\end{eqnarray*}

In the previous example we observed that the term with the second
derivative in brackets appears as a result of approximation
of the first derivative using forward differencing. This observation
leads us to the following finite--difference approximation of the
given equation, which has the above first differential approximation:
\begin{eqnarray*}
   \frac{y_{\alpha+1}-y_\alpha}{x_{\alpha+1}-x_\alpha}
 - (x_\alpha + 1) y_\alpha^2 - y_\alpha
 +(x_{\alpha+1}-x_\alpha)
  \left(
       - \frac{(x_\alpha+1)^2}{x_\alpha}
         y_\alpha^2 (1+x_\alpha y_\alpha)
       - \frac{y_\alpha}{x_\alpha^2}
       + C(x_\alpha) y_\alpha (1+x_\alpha y_\alpha)
  \right)
= 0.
\end{eqnarray*}

As the above consideration demonstrates, adding the last term to the
finite--difference approximation results in the considered group becoming
stable.

\appendix

\section{Calculations from the proof of Lemma 1
} \label{appendix:1}

Derivation of equation (\ref{eq:lemma_1}):
\begin{eqnarray*}
& &
  u_{\alpha+l+1}^{(J)} = u_{\alpha}^{(J)}
                     + (u_{\alpha+l+1}^{(J)}-u_{\alpha+l}^{(J)})
                     + (u_{\alpha+l  }^{(J)}-u_{\alpha  }^{(J)}) \\
&=&
  u_{\alpha}^{(J)} +
  \sum_{k=1}^{p  }  \frac{\varepsilon^k h_{\alpha+l}^k}{k!}
     u_{\alpha+l}^{(J+k)} +
  \sum_{k=1}^{p-J}  \frac{\varepsilon^k h_{\alpha+l,\alpha}^k}{k!}
     u_{\alpha}^{(J+k)}
    + o(\varepsilon^{p-J}) \\
&=&
  u_{\alpha}^{(J)} +
  \sum_{k=1}^{p-J}  \frac{\varepsilon^k}{k!}
                 \left[
                    h_{\alpha+l      }^k u_{\alpha+l}^{(J+k)}
                   +h_{\alpha+l,\alpha}^k u_{\alpha  }^{(J+k)}
                 \right] + o(\varepsilon^{p-J}) \\
&=&
  u_{\alpha}^{(J)} +
  \sum_{k=1}^{p-J}  \frac{\varepsilon^k}{k!}
                  \left[
                    h_{\alpha+l      }^k \left\{
                                              \sum_{m=0}^{p-J-k}
                                              \frac{\varepsilon^m
                                                    h_{\alpha+l,\alpha}^m
                                                   }{m!}
                                              u_{\alpha}^{(J+k+m)}
                                        \right\}
                   +h_{\alpha+l,\alpha}^k u_{\alpha  }^{(J+k)}
                  \right] + o(\varepsilon^{p-J}) \\
&=&
  \sum_{k=1}^{p-J}\sum_{m=0}^{p-J-k}
  \frac{\varepsilon^{k+m}}{k!m!}
  h_{\alpha+l}^k h_{\alpha+l,\alpha}^m u_{\alpha}^{(J+k+m)}
 +\sum_{k=0}^{p-J}
  \frac{\varepsilon^k}{k!} h_{\alpha+l,\alpha}^k u_{\alpha  }^{(J+k)}
 + o(\varepsilon^{p-J}) \\
&=&
  \sum_{k=1}^{p-J}\sum_{i=k}^{p-J}
  \frac{\varepsilon^{i}}{k!(i-k)!}
  h_{\alpha+l}^k h_{\alpha+l,\alpha}^{(i-k)} u_{\alpha}^{(J+i)}
 +\sum_{i=0}^{p-J}
  \frac{\varepsilon^i}{i!} h_{\alpha+l,\alpha}^i u_{\alpha  }^{(J+i)}
 + o(\varepsilon^{p-J}) \\
&=&
  \sum_{k=0}^{p-J}\sum_{i=k}^{p-J}
  \frac{\varepsilon^{i}}{k!(i-k)!}
  h_{\alpha+l}^k h_{\alpha+l,\alpha}^{i-k} u_{\alpha}^{(J+i)}
 + o(\varepsilon^{p-J}) \\
&=&
  \sum_{i=0}^{p-J}
  \frac{\varepsilon^{i}}{i!} u_{\alpha}^{(J+i)}
  \sum_{k=0}^{i}
  \frac{i!}{k!(i-k)!} h_{\alpha+l}^k h_{\alpha+l,\alpha}^{i-k}
  + o(\varepsilon^{p-J})
 =
  u_{\alpha}^{(J)} +
  \sum_{i=1}^{p-J}
  \frac{\varepsilon^{i}}{i!}
  h_{\alpha+l+1,\alpha}^i u_{\alpha}^{(J+i)}
  + o(\varepsilon^{p-J}),
\end{eqnarray*}

To prove (\ref{eq:lemma_2})
use (\ref{eq:step_size}), (\ref{eq:fd_derivatives}),
(\ref{eq:zero_derivative}) and the induction hypothesis:
\begin{eqnarray*}
  u_{\alpha-1}^{(J_0)} &=& u_{\alpha}^{(J_0)}
 -\sum_{k=1}^{p-J_0} \frac{\varepsilon^k h_{\alpha-1}^k}{k!}
  u_{\alpha-1}^{(J_0+k)} + o(\varepsilon^{p-J_0}) \\
&=& u_{\alpha}^{(J_0)}
 -\sum_{k=1}^{p-J_0} \frac{\varepsilon^k h_{\alpha-1}^k}{k!}
  \left[
        \sum_{m=0}^{p-J_0} \frac{\varepsilon^m (-h_{\alpha-1})^m}{m!}
        u_{\alpha}^{(J_0+k+m)}
  \right]
 + o(\varepsilon^{p-J_0})  \\
&=& u_{\alpha}^{(J_0)}
 -\sum_{k=1}^{p-J_0} \sum_{m=0}^{p-J_0}
  \varepsilon^{(k+m)}(-1)^m
  \frac{h_{\alpha-1}^{(k+m)}}{k!m!}
  u_{\alpha}^{(J_0+k+m)}
 + o(\varepsilon^{p-J_0})  \\
&=&u_{\alpha}^{(J_0)}
 -\sum_{k=1}^{p-J_0} \sum_{i=k}^{p-J_0}
  \varepsilon^{i}(-1)^{i-k}
  \frac{h_{\alpha-1}^{i}}{k!(i-k)!}
  u_{\alpha}^{(J_0+i)}
 + o(\varepsilon^{p-J_0}) \\
&=&u_{\alpha}^{(J_0)}
 -\sum_{i=k}^{p-J_0}
  \varepsilon^{i}(-1)^{i}
  \frac{h_{\alpha-1}^{i}}{i!}
  u_{\alpha}^{(J_0+i)}
  \left[
    \sum_{k=1}^{i}
    (-1)^{k} \frac{i!}{k!(i-k)!}
  \right]
  + o(\varepsilon^{p-J_0}) \\
&=&u_{\alpha}^{(J_0)}
 -\sum_{i=k}^{p-J_0}
  \varepsilon^{i}(-1)^{i}
  \frac{h_{\alpha-1}^{i}}{i!}
  u_{\alpha}^{(J_0+i)}
  \left[
    (1-1)^i-1
  \right]
  + o(\varepsilon^{p-J_0}) \\
&=&u_{\alpha}^{(J_0)}
 +\sum_{i=k}^{p-J_0}
  \varepsilon^{i}
  \frac{(-h_{\alpha-1})^{i}}{i!}
  u_{\alpha}^{(J_0+i)}
 + o(\varepsilon^{p-J_0}),
\end{eqnarray*}

Similarly, utilizing (\ref{eq:step_size}), (\ref{eq:fd_derivatives}),
(\ref{eq:zero_derivative})  and the induction hypothesis, we derive
(\ref{eq:lemma_3}):
\begin{eqnarray*}
  u_{\alpha+\beta_0-1}^{(J_0)}
&=& \sum_{k=0}^{p-J_0}
  \frac{\varepsilon^{k} (-h_{\alpha+\beta_0-1})^k}{k!}
  u_{\alpha+\beta_0}^{(J_0+k)}
  + o(\varepsilon^{p-J_0})  \\
&=& \sum_{k=0}^{p-J_0}
  \frac{\varepsilon^{k} (-h_{\alpha+\beta_0-1})^k}{k!}
  \left[
    \sum_{m=0}^{p-J_0-k}
    \frac{\varepsilon^m h_{\alpha+\beta_0,\alpha}^m}{m!}
    u_{\alpha}^{(J_0+k+m)}
  \right]
  + o(\varepsilon^{p-J_0})\\
&=& \sum_{k=0}^{p-J_0} \sum_{i=k}^{p-J_0}
  \frac{\varepsilon^{k}}{k!} \;\;
  (-h_{\alpha+\beta_0-1})^k \;\; h_{\alpha+\beta_0,\alpha}^{i-k}
  \;\; u_{\alpha}^{(J_0+i)}
+ o(\varepsilon^{p-J_0})  \\
&=& \sum_{i=0}^{p-J_0}
  \frac{\varepsilon^{i}}{i!}
  u_{\alpha}^{(J_0+i)}
  \left[
    \sum_{k=0}^{i}
    \frac{i!}{k!(i-k)!}
    (-h_{\alpha+\beta_0-1})^k
    (h_{\alpha+\beta_0,\alpha}^{i-k})
  \right]
+ o(\varepsilon^{p-J_0}) \\
&=& \sum_{i=0}^{p-J_0}
  \frac{\varepsilon^{i}}{i!}
  u_{\alpha}^{(J_0+i)}
  (
   -h_{\alpha+\beta_0-1}
   +h_{\alpha+\beta_0,\alpha}
  )^i
+ o(\varepsilon^{p-J_0}) \\
&=& u_{\alpha}^{(J_0)}
 +\sum_{i=1}^{p-J_0}
  \frac{\varepsilon^{i}}{i!}
  u_{\alpha}^{(J_0+i)}
  h_{\alpha+\beta_0-1,\alpha}^i
+ o(\varepsilon^{p-J_0}),
\end{eqnarray*}

\section{Calculations from the proof of Lemma 2
} \label{appendix:2}

To obtain equation (\ref{eq:lemma2_1}), use the notations
introduced in the proof of Lemma~\ref{lemma:expansion_of_a_function}
and write:
\begin{eqnarray}
   \left(
           \sum_{l=1}^{p} (\varepsilon h_\alpha)^l L_l
   \right)^n
=  \sum_{l=n}^{p} (\varepsilon h_\alpha)^l L^{(l)}_{(n)}
   + o(\varepsilon^{p}).  \label{eq:appendix:2_1}
\end{eqnarray}
Using this relation and changing the order of summation,
write
\begin{eqnarray*}
     \xi(x_{\alpha+1}, u_{\alpha+1})
&=&  \sum_{k=0}^{p} \sum_{i=0}^{k} \frac{1}{k!}
     \left\{
          C_{i}^{k} \frac{\partial^k \xi(x_\alpha,u_\alpha)}
                         {\partial x^i \partial u^{k-i}}
          (x_{\alpha+1} - x_{\alpha})^i
          (u_{\alpha+1} - u_{\alpha})^{k-i}
     \right\} + o(\varepsilon^{p})                    \\
&=&  \sum_{k=0}^{p} \sum_{i=0}^{k} \frac{1}{k!}
     \left\{
          C_{i}^{k} \frac{\partial^k \xi(x_\alpha,u_\alpha)}
                         {\partial x^i \partial u^{k-i}}
          (\varepsilon h_\alpha)^i
          \left(
                  \sum_{l=1}^{p} (\varepsilon h_\alpha)^l
                  \frac{u_\alpha^{(l)}}{l!}
          \right)^{k-i}
     \right\} + o(\varepsilon^{p})                   \\
&=& \sum_{k=0}^{p} \sum_{i=0}^{k} \sum_{l=k-i}^{p} \frac{1}{k!}
    C_{i}^{k} \frac{\partial^k \xi(x_\alpha,u_\alpha)}
                   {\partial x^i \partial u^{k-i}}
                   (\varepsilon h_\alpha)^{i+l} L_{(k-i)}^{(l)}
              + o(\varepsilon^{p})                   \\
&=& \sum_{m=0}^{p} \frac{(\varepsilon h_\alpha)^m}{m!}
    \left\{
            \sum_{i=0}^{m} \sum_{k=i}^{m}
            \frac{m!}{k!} C^i_k
            \frac{\partial^k \xi(x_\alpha,u_\alpha)}
                 {\partial x^i \partial u^{k-i}}
            L_{(k-i)}^{(m-i)}
    \right\} + o(\varepsilon^{p}) \\
&=& \sum_{m=0}^{p} \frac{(\varepsilon h_\alpha)^m}{m!}
    \left[
          \sum_{i=0}^{m}
          C_m^i \left(
                       \sum_{n=0}^{m-i}
                       \frac{(m-i)!}{n!} L^{(m-i)}_{(n)} \partial_u^n
                \right)
                \partial_x^i \xi(x_{\alpha},u_{\alpha})
    \right]
    + o(\varepsilon^p).
\end{eqnarray*}

Prove now equation (\ref{eq:lemma2_3}) with $l=k+1$
using the fact that the same equation is valid for $l=k$.
First notice that, as follows from (\ref{eq:appendix:2_1}),
$L^{(l)}_{(n)}$ can be expressed as follows:
\begin{eqnarray}
   L^{(l)}_{(n)} = \left.
                   \left(
                           \frac{1}{l!} \frac{d^l}{d a^l}
                           \left(
                                   \sum_{i=1}^p \frac{a^i}{i!}u^{(i)}
                           \right)^n
                   \right)
                   \right|_{a=0}. \label{eq:appendix:2_2}
\end{eqnarray}
Having applied the operator $\sum_J u^{(J+1)} \partial_{u^{(J)}}$
to both sides of equation (\ref{eq:lemma2_3}) with $l=k$,
use (\ref{eq:appendix:2_2}) and the expressions for $L^{(k)}_{(k)}$
and $L^{(k)}_{(0)}$ to transform the left--hand side
as follows (the arguments of $\xi$ are omitted for brevity):
\begin{eqnarray*}
& &  \sum_J u^{(J+1)} \partial_{u^{(J)}}
  \left\{
        \sum_{n=0}^{k}
        \frac{k!}{n!} L^{(k)}_{(n)} \partial_u^n \xi
  \right\}
 = \sum_{n=0}^{k} \frac{k!}{n!} L^{(k)}_{(n)}
      u^{(1)}\partial_u^{n+1} \xi
   +\sum_{n=0}^{k}
        \frac{k!}{n!}
        \left(
               \sum_J u^{(J+1)} \partial_{u^{(J)}} L^{(k)}_{(n)}
        \right) \partial_u^n \xi \\
&=& \sum_{n=1}^{k} \frac{k!}{n!}
    \left\{
            L^{(k)}_{(n-1)} n u^{(1)}
          + \sum_J u^{(J+1)} \partial_{u^{(J)}} L^{(k)}_{(n)}
    \right\} \partial_u^n \xi + (u^{(1)})^{k+1} \partial_u^{n+1} \xi  \\
&=& \sum_{n=1}^{k} \frac{k!}{n!} \frac{n}{k!} \frac{d^k}{d a^k}
    \left.
    \left\{
           \left(
                  \sum_{i=1}^p \frac{a^i}{i!} u^{(i)}
           \right)^{n-1}
           \sum_{i=0}^p \frac{a^i}{i!} u^{(i+1)}
    \right\}
    \right|_{a=0} \partial_u^n \xi + (u^{(1)})^{k+1} \partial_u^{n+1} \xi \\
&=&\sum_{n=1}^{k} \frac{k!}{n!} \frac{n}{k!} \frac{d^k}{d a^k}
    \left.
    \left\{
           \left(
                  \sum_{i=1}^p \frac{a^i}{i!} u^{(i)}
           \right)^{n-1}
           \sum_{i=0}^p \frac{a^{i-1}}{(i-1)!} u^{(i)}
    \right\}
    \right|_{a=0} \partial_u^n \xi + (u^{(1)})^{k+1} \partial_u^{n+1} \xi \\
&=&\sum_{n=1}^{k} \frac{k!}{n!} \frac{n}{k!} \frac{d^k}{d a^k}
    \left.
    \left\{
           \left(
                  \sum_{i=1}^p \frac{a^i}{i!} u^{(i)}
           \right)^{n-1}
           \sum_{i=0}^p \frac{a^{i-1}}{(i-1)!} u^{(i)}
    \right\}
    \right|_{a=0} \partial_u^n \xi + (u^{(1)})^{k+1} \partial_u^{n+1} \xi \\
&=&\sum_{n=1}^{k} \frac{(k+1)!}{n!} \frac{1}{(k+1)!} \frac{d^{k+1}}{d a^{k+1}}
    \left.
    \left\{
           \left(
                  \sum_{i=1}^p \frac{a^i}{i!} u^{(i)}
           \right)^n
    \right\}
    \right|_{a=0} \partial_u^n \xi + (u^{(1)})^{k+1} \partial_u^{n+1} \xi \\
&=&  \sum_{n=1}^{k} \frac{(k+1)!}{n!} L^{(k+1)}_{(n)} \partial_u^n \xi
  =  \sum_{n=0}^{k} \frac{(k+1)!}{n!} L^{(k+1)}_{(n)} \partial_u^n \xi,
\end{eqnarray*}
and therefore, equation (\ref{eq:lemma2_3}) with $l=k+1$
holds.

\end{document}